\documentclass[aps,paper,twocolumn]{revtex4}
\usepackage{amsmath}
\usepackage{amsfonts}
\usepackage{color}
\usepackage{amssymb}
\usepackage{graphicx}                     
\usepackage{epstopdf}

\begin{document}               

\title{Temporal disorder does not forbid 
discontinuous absorbing phase transitions in low dimensional systems}
\author{M. M. de Oliveira$^{1}$ and C. E. Fiore$^2$}
\address{
$^1$Departamento de F\'{\i}sica e Matem\'atica,
CAP, Universidade Federal de S\~ao Jo\~ao del Rei,
Ouro Branco-MG, 36420-000 Brazil, \\
$^2$ Instituto de F\'isica, Universidade de S\~ao Paulo, 
S\~ao Paulo-SP,  05314-970, Brazil
}

\date{\today}

\begin{abstract}
Recent papers have shown that spatial (quenched) 
disorder can suppress  discontinuous absorbing phase transitions.
Conversely, the scenario for temporal disorder  is still unknown.
In order to shed some light in this direction, we investigate its effect
in three different two dimensional models which are known to exhibit  
 discontinuous absorbing phase transitions.
The temporal disorder is introduced 
by allowing the control parameter to be time dependent $p\rightarrow p(t)$, 
either varying as a uniform distribution with mean ${\bar p}$ and variance $\sigma$ or as a bimodal distribution, fluctuating between a value $p$ and a value $p_l \ll p$. 
In contrast to  spatial disorder, 
our numerical results strongly suggest that such uncorrelated temporal
disorder does not forbid the existence of a discontinuous 
absorbing phase transition.
We find that all cases  are characterized by  
behaviors similar to their pure (without disorder) counterparts,
including bistability around the coexistence point and 
common finite size scaling behavior with the inverse of the system volume, as 
recently proposed in Phys. Rev. E. {\bf 92}, 062126 (2015). We also observe that 
temporal disorder does not induce temporal Griffiths phases around discontinuous phase transitions,
at least for $d=2$. 

\end{abstract}

\maketitle
\section{Introduction}
 Nonequilibrium phase transitions are considered a key feature of a countless number of phenomena, such as 
magnetic systems,  biological and ecological models, water-like anomalies,
and many others \cite{marr99,hinrichsen,odor04,henkel}. 
Recently, a considerable interest has been
devoted to the inclusion of more realistic 
ingredients in order to describe (or mimick) the effects
 of impurities or  external 
fluctuations, as well as their effects
in the phase transition \cite{buendia, bustos, liu, buono, paula,voronoi}.

Commonly, these ingredients are  introduced
by allowing the control parameter to assume distinct
values in  space and/or  time. 
The former case, regarded as quenched
disorder, affects drastically the phase transitions, 
leading to the existence of a new
 universality classes and local regions in the absorbing phases, 
characterized by large activities with slow decays towards  
extinction. These rare
regions  typically arise  
when the activation  rate  $\lambda$ lies between the clean value 
 $\lambda_c^0$ (without disorder) and the dirty (disordered) 
critical point $\lambda_c$; ie,  $\lambda_c^0<\lambda<\lambda_c$.
Moreover, in these regions the system may exhibit non-universal exponents toward
 full extinction \cite{igloi,oliveira,vojta}. 
Heuristically,  the Harris criterion 
\cite{harris}  establishes that quenched
disorder  is a relevant perturbation
if $d\nu_{\perp} <2$, where $d$ the system
dimensionality and $\nu_{\perp}$  is the spatial correlation
length exponent. 
For models belonging to
the directed percolation (DP) universality class  $\nu_{\perp}=1.096854(4), 0.734(4)$ and $0.581(5)$ 
in $d=1,2$ and $3$, respectively. Consequently,  the Harris criterion indicates
that spatial disorder 
is a relevant perturbation for continuous absorbing phase transitions in all dimensions.

Conversely, the Imry-Ma
 \cite{imry} and Aizenman-Wehl \cite{wehl} criteria establish that  
quenched disorder suppresses the phase coexistence
in equilibrium systems for $d \le 2$. 
Afterwards, it was shown  \cite{hovi,corte,hoenicke} that 
the discontinuous   transition  in the 
Ziff-Gulari-Barshad 
(ZGB) model becomes  continuous when the disorder strength is large enough. 
More recently, Villa-Mart\'in et al. \cite{paula} have suggested that 
the Imry-Ma-Aizenman-Wehl conjecture should be extended for 
discontinuous absorbing phase transitions for $d \le 2$, 
irrespective of the disorder magnitude. 

Although  less studied than  spatial disorder,  the 
influence the temporal disorder has also
been considered in some cases \cite{munoz2011,martinez,jensen96}.
In contrast to the quenched disorder, here 
the control parameter becomes time-dependent, 
resulting in a temporarily active (ordered)
as well as absorbing (disordered) phases,
whose effect of variability  becomes pronounced  at the 
emergence of the phase transition. 
In particular, the  available results 
have shown that temporal
disorder is a highly relevant perturbation \cite{vojta-hoyos}, 
suppressing  the DP phase
transitions in all dimension. For systems with up-down symmetry
they are relevant only for $d\ge 3$. 
{\em Temporal Griffiths phases} (TGPs),  a region in the active
phase characterized by  power-law spatial scaling and generic 
divergences of the susceptibility,
have also been reported for absorbing phase transitions 
\cite{munoz2011,vojta-hoyos,neto2,solano}, 
but not found in low dimensional systems
with up-down symmetry \cite{martinez}.
On the other hand, the effect of temporal disorder for
{\em discontinuous} absorbing phase transitions is still unknown.

In order to shed some light in this direction, here we investigate
the effects of temporal disorder in discontinuous
absorbing phase transition. Our study aims
to answer three fundamental questions: (i) is the occurrence of
phase coexistence forbidden under the presence of temporal disorder? (ii) if no, which changes does
it provoke with respect to the pure (without disorder) version? 
(iii) Does the temporal disorder induce temporal
Griffiths phases around these phase transitions?
These ideas will be tested in three models which are known to yield discontinuous absorbing phase transitions in two- and infinite-dimensional  systems, 
namely the ZGB model for CO oxidation \cite{zgb}
and two lattice versions of the second Schl\"ogl model  (SSM) \cite{schlogol72,oliveira}.
As we will show, in all cases the phase transition is characterized by a behavior similar to their pure (without disorder) counterparts,
including bistability around the coexistence point and
common finite size scaling behavior with the inverse of the system volume, as
recently proposed in \cite{martins-fiore}.
 
This paper is organized as follows:  In Sec. II we review the models studied 
and the simulation methods employed. Results and discussion are shown in Sec. III and conclusions are presented
in Sec. IV.

\section{Models and methods}

The SSM is  single-species autocatalytic reaction model defined
by the  reactions $2A \to 3A$ and $A\to 0$, 
which occurs with transition rates 
$1$ and $\alpha$, respectively. 
Such system displays a discontinuous phase transition that can be qualitatively 
reproduced under a mean-field treatment. 
The first reaction predicts
a particle growth following a quadratic 
dependence on the density, which makes 
 low-density (active) state unstable and thus,
a jump  to a nonzero  (large) density 
arises as the creation probability 
$1/(1+\alpha)$ increases to
a threshold value $\alpha_0=1/4$ \cite{marr99}.
Nonetheless, distinct works have claimed that  
these reaction rules are not sufficient to exhibit a discontinuity in a 
regular
lattice \cite{durret}. In particular, the system dimensionality 
and the geometrical constraint of requiring the 
presence of  a pair of adjacent particles surrounding an 
empty site (in order to fill the reaction $2A \to 3A$) 
are essential ingredients for the emergence of a phase coexistence 
\cite{fiore14,foot2}. 

Here, we consider two square lattice versions of the SSM. The first one (SSM1), 
proposed by Windus and Jensen \cite{jensen} and afterwards reconsidered  
in Ref. \cite{paula},  is defined as follows: 
A given particle  $i$  is 
 chosen (with equal probability) from a list of currently occupied sites
and is annihilated with probability  $p_a=\alpha/(1+\alpha)$. Or, with
probability  $(1-p_a)/4$, a nearest
neighbor site of $i$, called site $j$, is also chosen 
at random.  If
the site $j$ is empty, the particle $i$ will diffuse for it. 
If $j$ is filled by a particle, an offspring
will be  created at one of the neighboring sites of
$i$ and $j$ (chosen with equal possibility)
with probability $p_b$  provided it is empty; otherwise
nothing happens. The value  $p_{b} = 0.5$ has been considered 
to directly  compare our results with previous  
studies \cite{jensen,paula,martins-fiore}. 
After above dynamics, the time is incremented by  $1/N$, where $N$ is 
the number of occupied sites.

For the second version, SSM2, 
the  selection of particle $i$, its annihilation probability  
and the choice of the nearest
neighbor site  $j$ are identical to SS1. 
However, in the SSM2
when  a neighboring site $j$ is chosen, 
its  number of nearest neighbor occupied sites $nn$ will be evaluated. A new 
offspring will be created at $j$ with rate $nn/4$ provided $nn \ge 2$
and it is empty.   More specifically, if $nn=1$ no particle will be 
created in the vacant site. On the contrary, 
if $nn=2$, $3$ or $4$, the  creation will occur
with probability $nn/4$.

It is worth mentioning that in the SSM1, 
the discontinuous transition is caused by both the diffusion and the creation of offsprings in the presence of two particles. 
Conversely, in the SSM2 model it is caused by the creation of offsprings in the presence of at least two species.  


The third system we investigate is the ZGB model \cite{zgb}, 
which qualitatively reproduces some  features of the  
oxidation of carbon monoxide on a  catalytic
surface. The surface is modeled as a square lattice, 
in which each site can be empty ($*$), or occupied by an 
oxygen (O$_{ads}$) or a carbon monoxide (CO$_{ads}$).
It is summarized by the following reactions:
\begin{eqnarray}
\mbox{CO}_{gas}+*\to \mbox{CO}_{ads} \nonumber \\
\mbox{O}_{2 gas}+2* \to 2\mbox{O}_{ads} \nonumber \\
\mbox{CO}_{ads}+\mbox{O}_{ads}\to \mbox{CO}_{2}+ 2*. \nonumber
\end{eqnarray}
\noindent 
In practice, molecules of CO$_{gas}$ and O$_{2 gas}$ hit the surface with
complementary probabilities $Y$ and $(1-Y)$, respectively, at any time 
the chosen site is empty.  At the surface, O$_2$ molecule dissociates into two 
independent O atoms, each one occupying two adjacent empty 
sites. If a CO$_{ads}$O$_{ads}$ pair is placed at neighboring sites on
the surface, a CO$_{2}$ molecule will be formed, desorbing  instantaneously and leaving both sites empty. 
As in the SSM models, after the above dynamics
is implemented,  time is incremented by $1/N$ where $N$ is the total number of empty sites. 

By changing the parameter $Y$, the model exhibits two phase transitions 
between an active steady state and one of two absorbing (``poisoned'') states,  in which the 
surface is saturated either by O or by CO. The O-poisoned transition 
is found to be continuous. On the other hand, the CO-poisoned 
transition is discontinuous, and  in this work we will focus on this specific case.
 
For the SSMs, the order parameter $\phi$ is the system density $\rho$ and
the transitions take place at
 $\alpha_0=0.0824(1)$ (SSM1) \cite{martins-fiore,foot}
and $\alpha_0=0.2007(6)$ (SSM2) \cite{oliveira2}.
For the ZGB, $\phi$ is the density of CO and the transition occurs
at $Y_0=0.5250(6)$\cite{martins-fiore}.

The temporal disorder is introduced so that at  
each  time interval $t_i\le t\le t_i+\Delta t$, a generic control parameter 
$p$ assumes  a value extracted from a uniform
distribution with mean ${\bar p}$ and width $\sigma$. More specifically, 
$p$ is evaluated using the formula  $p={\bar p}+(2\xi-1)\sigma$,
where $\xi$ is a random number drawn at each time interval $\Delta t$ from the standard uniform distribution  in

 $ [0,1]$.
For the SSMs, ${\bar p}$ corresponds to 
the creation probability ${\bar p}=1-p_a=\frac{1}{1+\alpha}$,
with a similar formula holding for the ZGB model with $1-p_a$ replaced by $Y$. 

In order to locate the transition point and   the nature
of the phase transition, we consider three alternative procedures. 
First, we follow the time behavior of the order-parameter 
$\phi(t)$, starting from
a fully active initial configuration. 
In the active phase, it converges to a constant value,
signaling the 
permanent  creation and annihilation of particles. 
On the other hand,  $\phi(t)$ decays exponentially
toward  extinction (full poisoned state for the ZGB) in the absorbing phase.
In the case of a typical (without disorder) 
continuous phase transitions, the above regimes are separated
by a power-law decay   $\phi(t) \sim t^{-\theta}$, with
$\theta$ being the associated critical exponent. For the DP universality
class, $\theta=0.4505(10)$ in two dimensions \cite{henkel}. In the presence
of temporal disorder, the above critical behavior is replaced
by  $\phi(t) \sim (\ln t)^{-1}$ \cite{neto2,solano}.
Additionally,  one does not expect
 similar behaviors at the emergence of a discontinuous transition.

The coexistence point can be estimated through a 
threshold value ${\tilde \alpha}$, which separates the saturation toward a definite value from an exponential 
decay \cite{fiore14,fioresal,fioresal2}. 
Alternatively, a more reliable procedure is achieved by 
performing a finite-size analysis, as  
 recently proposed in  Ref.~\cite{martins-fiore}. According to it, the 
difference between the pseudo-transition point  $\alpha_L$ and 
the transition point $\alpha_0$ scales with  $L^{-2}$, where $L^2$ denotes
the system volume (in two dimensions)
The estimation of $\alpha_L$ can be done in  a variety of ways. 
For instance, as corresponding to the the peak of the system's order-parameter variance 
$\chi=L^{2}(\langle \phi^2\rangle-\langle \phi \rangle^2)$, or even
through the value in which the bimodal order parameter distribution presents
two equal areas \cite{martins-fiore}.
However, such scaling behavior is verified only 
by considering some kind of quasi-stationary (QS) ensemble, i.e. 
an ensemble of states accessed
by the original dynamics at long times {\it conditioned on survival}
(and restricted to those which are not trapped into an absorbing state). 
Here we employ an efficient numerical scheme 
given in Ref. \cite{qssim}, in which configurations are stored and gradually updated
 during the evolution of the stochastic process. 
Whenever  the transition to the absorbing state is imminent, the system is ``relocated'' to a
saved configuration. This accurately reproduces the results
from the much longer procedure of performing averages only
on samples that have not been trapped in the absorbing state at the end of their
respective runs.
The intensive quantities in a QS ensemble must converge to the
stationary ones when  $L \to \infty$. 

Finally, in the third procedure,  the mean survival time $\tau$ is considered for different system sizes.
According to Refs. \cite{paula,ref30}, the coexistence point is the separatrix of
an exponential growth of $\tau$ and an exponential increase until $L < L_c$
followed by a decreasing behavior  for $L>L_c$. Here, we shall also
quantify it, in order to compare with the pure (not disordered) cases. 

\section{Results and discussion} 

\subsection{Models in a square lattice}

\begin{figure}[h]
\centering
\includegraphics[scale=0.345]{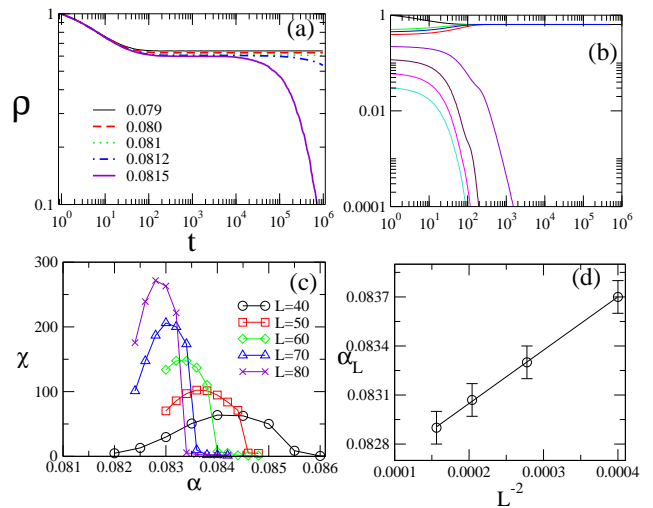}
\caption{({\bf Color online}): Results for the pure SSM1. 
Panel $(a)$ shows the time decay of $\rho(t)$ for 
$\rho(0)=1$ and distinct values of $\alpha$. Panel
$(b)$ shows the bistable behavior of $\rho(t)$ close to the
separatrix point ${\tilde \alpha} \sim 0.0815$ for distinct initial
densities  ranging
from $10^{-2}$ to $1$. 
Panel $(c)$ and $(d)$ shows the order parameters variance 
$\chi$ versus $\alpha$ and the value   $\alpha_L$ for which $\chi$ is maximum, vs $1/L^2$).}
\label{fig1}
\end{figure}

\begin{figure}[h]
\centering
\includegraphics[scale=0.345]{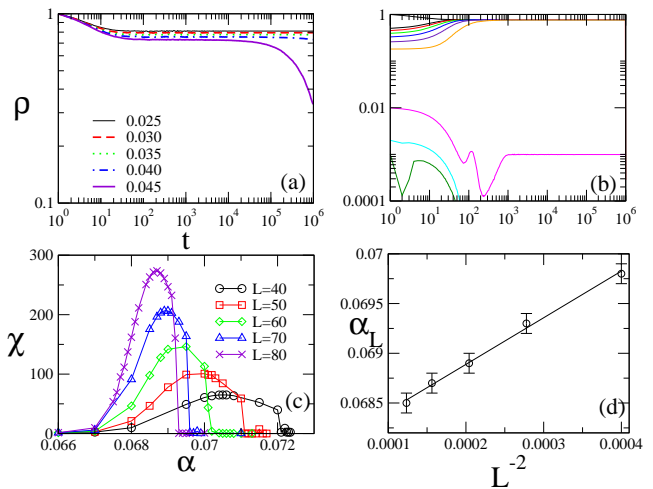}
\caption{({\bf Color online}): Results for $\sigma=0.15$. 
Panel $(a)$ shows the time decay of $\rho(t)$ for 
$\rho(0)=1$ and distinct values of $\alpha$. Panel
$(b)$ shows the bistable behavior of $\rho$ at ${\tilde \alpha} \sim 0.035$ for distinct initial
densities ranging from $10^{-2}$ to $1$. Panel $(c)$ and $(d)$ shows the order parameters variance 
$\chi$ versus $\alpha$ and the  value $\alpha_L$ for which $\chi$ is maximum, vs $1/L^2$).}
\label{fig3}
\end{figure}

The first analysis of the influence of temporal disorder 
is achieved by inspecting the time decay of the order parameter $\phi(t)$
starting from a fully active initial configuration for 
$t=0$.  For the SSM1, Figs.
\ref{fig1} and \ref{fig3} (panels $(a)$)
show $\rho(t)$ for $\rho(0)=1$ for the pure versions and for 
$\sigma=0.05$ (not shown) and $\sigma=0.15$, with $\Delta t=1$. 


In all cases there is a threshold value ${\tilde \alpha}$
separating  indefinite activity and exponential
decay toward the particle extinction. They 
are strongly dependent on $\sigma$ and occur 
at ${\tilde \alpha}=0.0812, 0.076$ (not shown)
and $0.035$ for the pure, $\sigma=0.05$ and $0.15$, respectively.
No indication of a power-law  have been verified nor
a behavior of type $\rho \sim (\ln t)^{-1}$. 
By repeating the above analysis for distinct initial 
configurations (panels $(b)$)
with distinct densities ($10^{-2} \le \rho(0)\le 1$), the curves
 converge  to two well defined  stationary states, 
with $\rho <<1$ and $\rho \sim \rho^*$,
signaling the bistability of active and absorbing phases, thus 
suggesting in all cases a first-order phase transition. 
For the pure, $\sigma=0.05$ and $0.15$,
$\rho^*$ read $0.637(2)$, $0.63(2)$ and $0.77(2)$ respectively.

Inspection of quasi-stationary properties for distinct $L$'s reveal
that the $\alpha_{L}$'s (panels $(c)$ and $(d)$), 
in which the  order parameter variance 
$\chi$ is maximum, scales with $1/L^{2}$ and gives 
$\alpha_{0} = 0.0824(2)$, $0.0823(2)$ (not shown) and $0.0680(2)$ 
for the pure,  $\sigma=0.05$ (not shown) and $0.15$, respectively. 
In particular for $L=100$, the peak in $\chi$ occurs at $0.0827(1)$, 
$0.0826(1)$ and $0.0684(1)$, respectively.
Therefore,  both previous  analyses  suggest
that temporal disorder does not forbid a discontinuous
phase transition. However, it  increases the  metastable region 
at the emergence of the phase coexistence, i.e. $\alpha_L-{\tilde \alpha}$ 
increases with $\sigma$. This feature shares similarities with
some procedures studied for characterizing the first-order transition 
in the ZGB and allied models close to the coexistence by taking different
initial configurations \cite{evans}.

\begin{figure}[h]
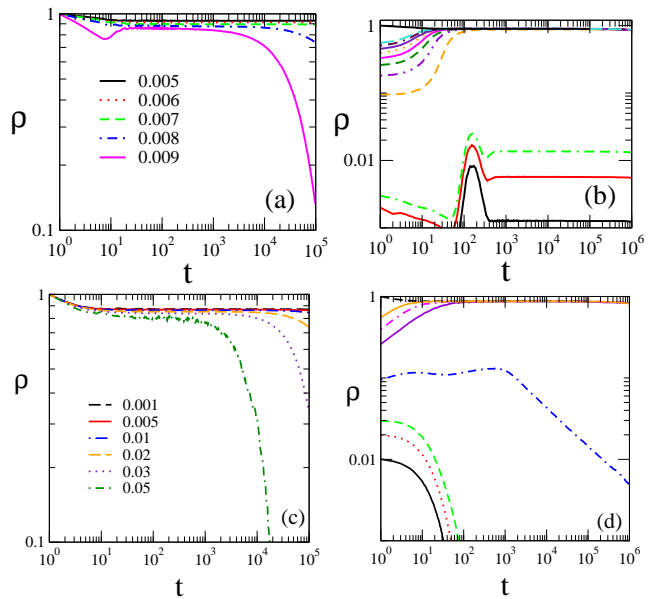

\centering
\includegraphics[scale=0.4]{jensen_str.eps}
\includegraphics[scale=0.35]{gra_thre_s0.4.eps}
\caption{({\bf Color online}) The SSM1 with
bimodal temporal disorder distribution. 
$(a)$  log-log time
decay of $\rho$ for distinct
$\alpha$'s and $\Delta t=6$.  $(b)$ Bistable behavior
of $\rho$ at $\alpha=0.007$ (very
close to the separatrix point ${\tilde \alpha} \sim 0.008$) for
distinct initial densities  ranging from $10^{-3}$ to $1$. Panel $(c)$  
shows the same analysis in $(a)$, but for the SSM2 with uniform
distribution for $\sigma=0.4$ and $\Delta t=1$.  In $(d)$, 
the bistable behavior
of $\rho$ at $\alpha=0.001$, 
close to the transition point ${\tilde \alpha} \sim 0.005$, for
distinct initial densities ranging from $10^{-3}$ to $1$.}
\label{fig2}
\end{figure}

An important point is that  above the transition  the number of points  
decrease substantially 
with increasing $\sigma$, revealing  a suppression (absence) of 
a phase transition for  $\sigma >0.22$, which is a rather small
disorder weight.
 In order to  strengthen (i.e., to increase) the influence of disorder in the SSM1,  we perform two changes. 
First we
increase the disorder duration $\Delta t$.  
Since no larger values of $\sigma$ are possible for this model, we change to a bimodal disorder distribution,
where, at each $\Delta t$, the creation probability is chosen from two values, $p=1/(1+\alpha)$ and 
$p_l=1/(1+20\alpha)$, with rates
$1-q$ and $q$, respectively. The results are presented
in Fig. \ref{fig2}(a), (b) for 
$q=0.2$, $L=200$ and $\Delta t=6$. 
Second, the analysis
of the SSM2 for a larger value $\sigma=0.4$ (with $\Delta t=1$) is 
considered. Since its pure
version yields a larger 
transition  point, it is possible to increase substantially the
value of $\sigma$ (in contrast to the SSM1).
These results  are presented in Fig. \ref{fig2}  (panels $(c)$ and $(d)$).

In both systems, the phase transitions obey the same pattern as the previous cases: 
separatrix points at ${\tilde \alpha}\sim 0.008$ (${\tilde \alpha}\sim 0.005$)
 and bistable behaviors of  $\rho(t)$ at 
the vicinity of the transition points [exemplified here for $\alpha=0.007$ (and $\alpha=0.001$)]. 
In both cases, the  ${\tilde \alpha}$'s are
very small,  highlighting the relevance
of disorder. 
As for the SSM1, the phase transition is suppressed for sufficiently large $\sigma$'s, whose
results reveal the absence of a phase transition  for $\sigma >0.4$. 

We now  turn our attention to the ZGB model. In Figs. 4 and Fig. 5 we show the 
results  for different disorder strengths, $\sigma = 0.05$ and
$\sigma=0.10$, respectively. In particular, we  have considered rather small disorder strengths, 
in order not to ``mix" both phase transitions. In both cases, 
 results similar to the SSMs have been  obtained.
Panels $(a)$ show once again the onset point ${\tilde Y}$  separating activity
from a exponential growth  toward a full  carbon monoxide poisoning.
The values of ${\tilde Y}$ decrease  by raising the disorder parameter $\sigma$,
and read ${\tilde Y}=0.527(1)$, $0.523(1)$, $0.516(1)$ and $0.500(2)$
for the pure, $\sigma=0.05$, $0.1$ and $0.2$ (not shown), respectively.
In addition, $Y_{L}$'s, obtained from the maximum of the order parameter variance 
$\chi$ scales with $1/L^{2}$ as seen in the pure version \cite{martins-fiore}. 
For the pure, $\sigma=0.05$, $0.1$ and $0.2$ (not shown) we obtain 
$Y_0=0.5253(3)$, $0.524(1)$, $0.520(1)$ and $0.509(2)$, respectively. 
Although less pronounced than for the previous
example, note that the difference $Y_0-{\tilde Y}$
increases with $\sigma$, reinforcing that 
disorder   increases the  spinodal region  around the phase coexistence.

\begin{figure}[!hbt]
\includegraphics[scale=0.35]{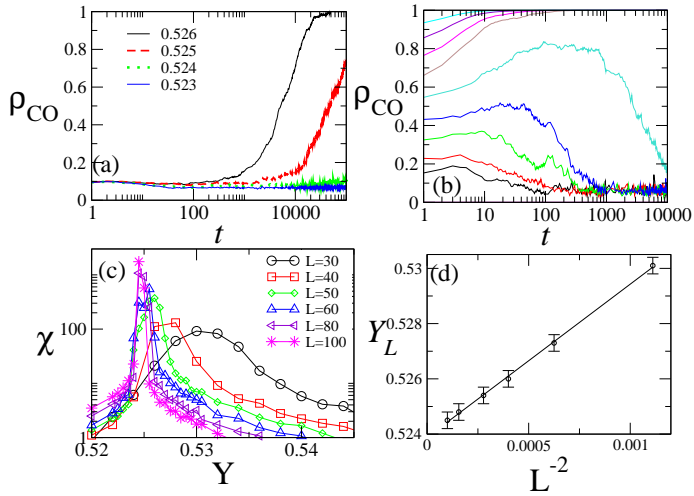}
\caption{({\bf Color online}):  Results for the ZGB model for $\sigma=0.05$. 
Panel $(a)$ shows the time decay of $\rho_{CO}$ for $\rho_{C0}(0)=0$ 
and distinct values of $Y$. Panel
$(b)$ shows the bistable behavior of $\rho_{CO}$  for $Y =  0.522$ for distinct initial
densities equi-spaced in the interval $[0.1,0.9]$ (Linear system size: $L=800$). Panel $(c)$ and $(d)$ shows the order parameters variance 
$\chi$ versus $Y$ and the  $Y_L$,
in which $\chi$ is maximum, vs $1/L^2$).}
\label{zgb2}
\end{figure}

\begin{figure}[!hbt]
\includegraphics[scale=0.35]{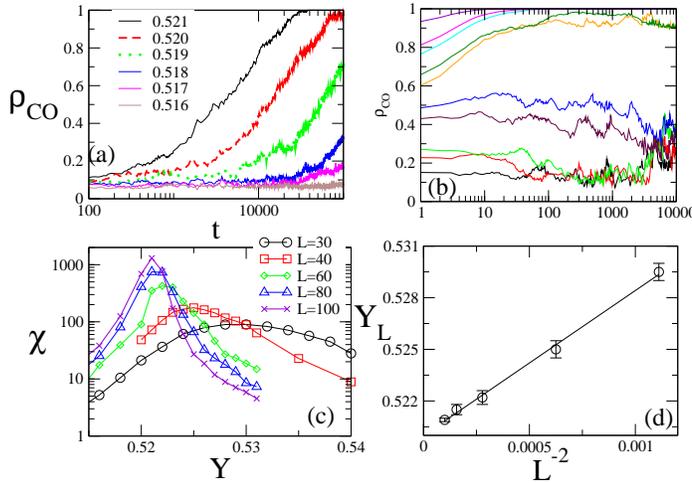}
\caption{({\bf Color online}): Results for the ZGB model for $\sigma=0.10$. 
Panel $(a)$ shows the time decay of $\rho_{CO}$ for $\rho_{C0}(0)=0$ 
and distinct values of $Y$. Panel
$(b)$ shows the bistable behavior of $\rho_{CO}$ for $Y =  0.520$ for distinct initial
densities equi-spaced in the interval $[0.1,0.9]$ (Linear system size: $L=800$). Panel $(c)$ and $(d)$ shows the order parameters variance 
$\chi$ versus $Y$ and the  $Y_L$,
in which $\chi$ is maximum, vs $1/L^2$).}
\label{zgb2}
\end{figure}


\begin{figure}[t]
\includegraphics[scale=0.35]{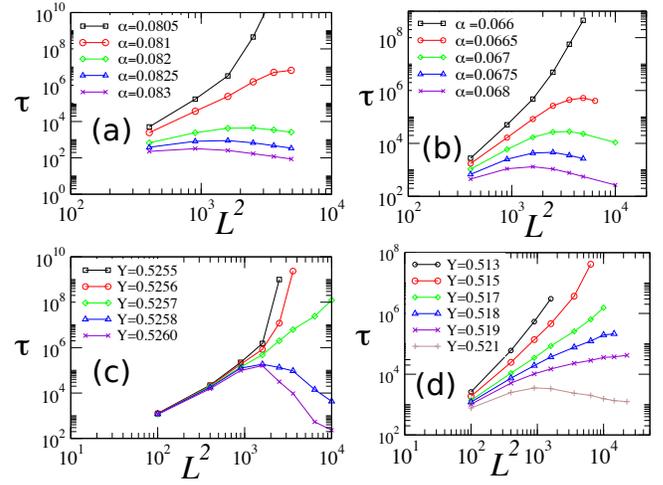}
\caption{({\bf Color online}): For the SSM1 
(ZGB) model,  panels $(a)$[$(c)$] and $(b)$[$(d)$] show the QS lifetime
for the pure and disordered versions, respectively. We 
take $\sigma=0.15$ and $0.1$ for the SSM1 
and ZGB models, respectively.} \label{tau}
\label{tau1}
\end{figure}

Fig. \ref{tau1}  shows the mean  lifetime of the QS state (defined as the time between
two absorbing attempts during the QS regime),
for the pure and disordered systems. 
We observe in all
cases the same behavior (in similarity with Ref. \cite{paula}): a threshold
value separating  exponential growth of $\tau$ up to a maximum  system size $L_c$, followed by a decrease
of $\tau$ for $L>L_c$.  For the pure cases, from such 
analysis  the coexistence points are located within the interval
 $0.0805<\alpha<0.081$ (SSM1)  and $0.5256<Y<0.5258$ (ZGB). In
the presence of temporal disorder,  they
are in the interval $0.066<\alpha<0.067$ (SSM1 for $\sigma=0.15$)
and $0.515<Y<0.518$ (ZGB for $\sigma=0.1$),  which
agrees with previous estimates  obtained from the maxima of $\chi$.
Thus the above findings suggest that in contrast  with critical 
transitions,  $\tau$ does not 
grow algebraically in a region within the active phase. These
results  are similar to those  obtained for  the generalized voter model
\cite{martinez}, suggesting
that TGPs do not manifest at discontinuous absorbing transitions, but only
at critical ones \cite{vojta-hoyos, neto2, solano}. However,  
this point still deserves further studies.

We close this section by remarking that the  active-CO poisoned  
transition exhibits a behavior consistent to a continuous transition
for $\sigma>0.3$ (not shown). Thus, in contrast to the SSMs (at least until $\sigma \le 0.4$)
numerical results indicate that the increase of $\sigma$ suppress the phase coexistence.

\subsection{Models in a complete graph} 

With the purpose of investigating the effects of temporal disorder  
in infinite-dimensional structures, the last analysis  considers a 
mean-field like description of  the above models, through  
 a complete graph (CG) treatment. 
In the CG approach, each site interacts with all others, so that  
an exact analysis is allowed. For the SSM, besides the  reactions
$A \rightarrow 0$ and $2A \rightarrow 3A$,  
one takes the coagulation process $2A \rightarrow A$
occurring with rate $\nu$ \cite{paula,dickman-vidigal02}. The
discontinuous transitions yield at the exact points
 $\alpha_0=1/(2\sqrt{\nu})$ and $Y_0=2/3$ \cite{zgbqs,dickman-vidigal02}, 
for the SSM and  ZGB, respectively. 
Due to the prohibition against C$-$CO 
occupying nearest-neighbor pairs, only one species (CO or O) may be present 
at any moment for the ZGB analysis. 
Let $\rho=\rho_{CO}-\rho_{O}$ with $\rho_{CO}$ and $\rho_O$ denoting the 
fraction of sites bearing a CO and O, 
respectively. This quantity allows to describe a system of 
$N$ sites completely by a single variable, with 
$\rho =-1$ representing the O-poisoned state and $\rho =1$ the 
CO-poisoned state 
(see more details in Ref. \cite{zgbqs}). 
In particular, we take $\nu=1$
for the SSM and in all cases 
the temporal disorder was introduced in a similar fashion way than in Sec. II.
\begin{figure}[t]
\includegraphics[scale=0.35]{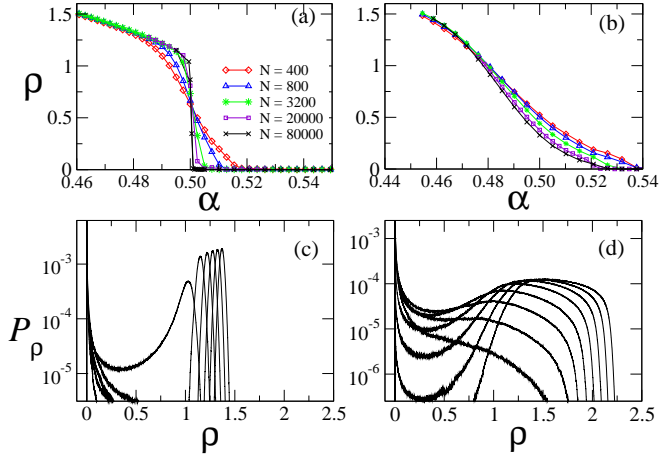}
\caption{({\bf Color online}): For  the SSM on a complete graph, 
the QS density $\rho$ for the pure model $(a)$ and  with temporal disorder 
strength $\sigma =0.1$ $(b)$. In $(c)$, the 
QS probability distributions for the pure model, with $\alpha$ ranging from $0.475$ to $0.525$. 
In $(d)$, the same as $(c)$ but for $\sigma=0.1$, and $\alpha$ ranging from $0.4750$ 
to $0.5625$. System size: $N=10000$ in $(c)$ and $(d)$. }
\label{schCG}
\end{figure} 

\begin{figure}[t]
\includegraphics[scale=0.35]{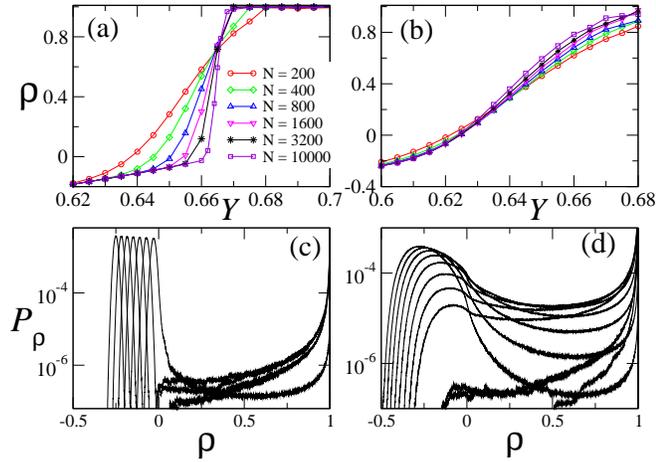}
\caption{({\bf Color online}): For  the ZGB model on a complete graph,
the QS order-parameter $\rho$ for the pure model $(a)$ and  with temporal disorder 
strength $\sigma=0.1$ $(b)$.  In $(c)$, the 
QS probability distributions for the pure model, with $Y$ ranging from $0.60$ to $0.70$. 
In $(d)$, the same as $(c)$ but for $\sigma=0.1$, and $Y$ ranging from $0.60$ 
to $0.70$. System size: $N=10000$ in panels $(c)$ and $(d)$. }
\label{zgbCG}
\end{figure} 

\begin{figure}[t]
\includegraphics[scale=0.35]{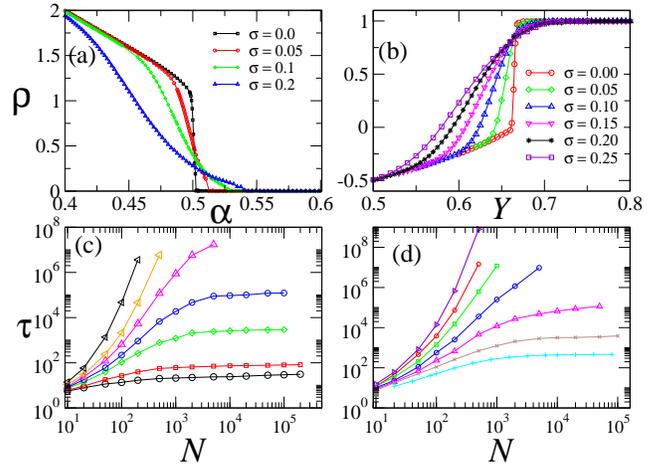}
\caption{({\bf Color online}): For the complete graph,
panels $(a)$ and $(b)$ show the QS order-parameter $\rho$ for the 
SSM and ZGB models,
respectively  for distinct $\sigma$'s and $N=10000$. Panel $(c)$ 
shows $\tau$ versus $N$ for the SSM model on 
a complete graph for $\sigma=0.1$ and $\alpha$ ranging from
$\alpha=0.458, 0.467, 0.472, 0.476, 0.481, 0.490$, and $0.500$ 
(from  top to bottom). In $(d)$, the same in $(c)$ but for the ZGB 
with $\sigma = 0.1$ and $Y$ ranging from 0.55 to 0.61 
(equi-spaced from  top to bottom).  }
\label{CG1}
\end{figure}

Our results for $\rho$ for the SSM and ZGB models are shown in panels $(a)$ 
of Figs.  \ref{schCG}  and \ref{zgbCG}, respectively. 
In both cases, the analysis in the complete graph predicts  behaviors which are similar to the numerical studies: the 
reduction of the active region, when
compared to their pure counterparts
and occurrence of bimodal probability
distributions [see e.g panels (c)-(d) in Figs. \ref{schCG} and \ref{zgbCG}]. 
In particular,  for disorder strength $\sigma=0.1$, the transition 
points are shifted from $\alpha_0=0.5$ 
to $\alpha_0=0.526$ (SSM), 
and from $Y_0=2/3$ to $Y_0=0.635$ (ZGB). Thus, the inclusion
of low disorder maintains the phase coexistence. However, by 
increasing $\sigma$ the active phase
peaks become broader, suggesting the appearance  of a continuous transition 
as shown in Fig. \ref{CG1} (a) and (b). Despite this,  there are 
some differences
when compared to their low
dimensional counterparts. There is a region
in the active phase (see e.g Fig. \ref{CG1} (c) and (d)),   
 in which  $\tau$  grows slower than exponential, 
and then it saturates at a finite value. This  behavior is
related to the abrupt transition that occurs when the noise takes the 
control parameter to a value that drives the system to
the absorbing state.
Since configurations with intermediary densities are unstable in these
 systems, one observes a bimodal QS probability distribution
in this region.  
This behavior is remarkably 
distinct from TGPs, in which $\tau$ increases algebraically 
with the system size $L$ and it has been observed 
only in continuous (absorbing) phase  transitions.

\section{Conclusions}

We studied the influence of temporal disorder in the
context of discontinuous absorbing phase transitions. We investigated extensively three models
 by means of distinct numerical procedures.  Our results strongly suggest  that in contrast
 to the spatial disorder, discontinuous absorbing transitions are not forbidden by the presence of 
temporal disorder in low dimensional systems. In particular,
the behavior of quantities are similar to their pure 
counterparts. However, the temporal disorder increases the metastable region close
to phase coexistence.  

Our results also suggest the absence of 
temporal Griffiths phases (TGPs). 
Some remarks over their existence  are in order:
Earlier results for different systems have shown that
the inclusion of temporal disorder does not
necessarily lead to the presence of TGPs \cite{martinez}. 
Although it suppresses the DP universality class in all dimensions,
the appearance of TGPs depend on $\sigma$ 
and/or $\Delta t$ \cite{neto2,solano}. Similar conclusions
continue to be  valid for  distinct up-down
systems, in which only for $d\ge 3$ TGPs  
are observed. Recent results for a one-dimensional example \cite{fiore17} confirm 
the  absence of TGPs 
 when the phase transition is discontinuous.

For the complete graph versions 
($d \rightarrow \infty$), we observe the maintenance   
of the phase coexistence for small disorder. However, in contrast
to the lattice versions, 
there is  a region in the active phase in which the lifetime 
 grows slower than exponential and then 
saturate at a finite value. 

It is worth emphasizing that our results do not exclude 
a discontinuous transition becoming continuous  from a disorder 
threshold $\sigma_c$.
Except for  the SSMs, in which the transition points decrease substantially as 
$\sigma$ increases,  
results for the ZGB model indicate  
the suppression of phase coexistence
 for $\sigma>0.3$. Again, the CG 
approach and the above mentioned one-dimensional 
case also reveal  similar trends. This last case shows that  the 
crossover to the criticality is also followed by 
appearance of TGPs within the active phase \cite{fiore17}.

Possible extensions of this work include the study of the effect of 
temporal correlated disorder and the more general case of spatio-temporal disorder, i.e. 
how the discontinuous phase transition is affected by an external perturbation that 
fluctuates in both space and time \cite{dickman-vojta}.
Both cases appear
to be of particular interest in the context
of ecosystems, where the effects of noise on the
extinction of a population
due to environmental changes have been attracting considerable
attention recently \cite{meerson}. 
Also, extension of both models for larger dimensions 
are intended to be investigated, in order to confirm the above hypotheses.

\section*{ACKNOWLEDGMENT}
We acknowledge Gabriel T. Landi and 
J. A. Hoyos for fruitful discussions.
The financial supports from CNPq and FAPESP, under grants 15/04451-2
and 307620/2015-8, are also acknowledged.

\bibliographystyle{apsrev}

\begin{thebibliography}{100}
\bibitem{foot2} As exemplified in distinct 
examples in \cite{fiore14}, whenever the $2A \to 3A$ is                   
replaced by $A\to 2A$, the transition becomes continuous.
\bibitem{foot}In fact,
our estimate is   different to the estimate $\alpha_c=0.0742(1)$
\cite{martins-fiore}, as a result of a slightly 
different dynamics implementation.
\end{thebibliography}

\end{document}